\documentclass[12pt]{article}

\usepackage{amssymb}
\usepackage{amsmath}
\usepackage{amscd}
\usepackage{latexsym}
\usepackage{graphicx}

\usepackage{cite}

\newcommand{\be}{\begin{equation}}
\newcommand{\ee}{\end{equation}}
\newcommand{\Dlt}{\Delta}
\newcommand{\dlt}{\delta}
\newcommand{\prt}{\partial}
\newcommand{\br}{{\bf r}}
\newcommand{\bk}{{\bf k}}
\newcommand{\bfe}{{\bf e}}
\newcommand{\ba}{{\bf a}}
\newcommand{\bp}{{\bf p}}
\newcommand{\bu}{{\bf u}}
\newcommand{\bt}{\beta}

\newcommand{\ep}{\varepsilon}
\newcommand{\al}{\alpha}
\newcommand{\ra}{\rightarrow}

\newcommand{\gm}{\gamma}
\newcommand{\om}{\omega}

\newcommand{\Gm}{\Gamma}
\newcommand{\dgr}{\dagger}

\newcommand{\rgl}{\rangle}
\newcommand{\lgl}{\langle}

\begin{document}

\begin{center}

{\Large{\bf Instability of insulating states in optical lattices due
to collective phonon excitations} \\ [5mm]

V.I. Yukalov$^{1,2}$ and K. Ziegler$^{1}$ } \\ [3mm]

{\it
$^1$Institut f\"{u}r Physik, Universit\"{a}t Augsburg, \\
D-86135 Augsburg, Germany \\ [3mm]

$^2$Bogolubov Laboratory of Theoretical Physics, \\
Joint Institute for Nuclear Research, Dubna 141980, Russia }

\end{center}

\vskip 5cm

\begin{abstract}

The role of collective phonon excitations on the properties of cold atoms 
in optical lattices is investigated. These phonon excitations are collective 
excitations, whose appearance is caused by intersite atomic interactions
correlating the atoms, and they do not arise without such interactions.
These collective excitations should not be confused with lattice vibrations
produced by an external force. No such a force is assumed. But the considered
phonons are purely self-organized collective excitations, characterizing  
atomic oscillations around lattice sites, due to intersite atomic interactions.   
It is shown that these excitations can essentially influence the possibility 
of atoms to be localized. The states that would be insulating in the absence 
of phonon excitations can become delocalized when these excitations are taken 
into account. This concerns long-range as well as local atomic interactions. 
To characterize the region of stability, the Lindemann criterion is used. 

\end{abstract}

\vskip 1cm


\newpage

\section{Introduction}

Cold atoms in optical lattices are usually considered in the frame of the
Hubbard model (see, e.g., reviews \cite{Morsh_1,Moseley_2,Bloch_3,Yukalov_4}).
The periodic potential of an optical lattice is imposed by external laser
beams. This potential is fixed in space, prescribing a lattice formed by
the lattice sites ${\bf a}_i$. We do not assume the existence of external 
fields that would move the lattice.  

If atomic interactions between different sites are neglected, the individual 
properties of atoms in each potential well are completely prescribed by the 
given optical lattice. However, as soon as the intersite atomic interactions 
are taken into account, the low-energy positions of the atoms are not exactly
those of the potential minima of the optical lattice. In other words, an atom 
in a potential well experiences an oscillational motion due to the interaction 
with the other atoms. These oscillations, due to the interaction effect, can 
be characterized as collective phonon excitations. 

It is these collective excitations that are considered in our paper. 
As is well known, collective excitations in many cases can essentially influence 
the system stability. Our aim is to study the role of such phonon excitations
for atoms in optical lattices. We use the standard method of taking into account 
collective excitations, by considering small deviations from equilibrium values. 

It is important to stress the necessity of intersite atomic interactions,
without which phonon excitations cannot exist. The situations of externally 
shaking the lattice and of the existence of self-organized collective excitations 
in a system of correlated atoms are principally different and should not be 
confused. In the former case, vibrations should be produced by an external field 
and would exist without any intersite atomic interactions. But in the latter case, 
there is no external shaking field and the collective excitations do not exist 
in the absence of intersite atomic interactions. 

It is the aim of the present paper to study the properties of insulating atomic
states in optical lattices, taking into account the arising phonon excitations
and their influence on the stability of the insulating states.  

It turns out that phonon excitations can essentially influence the properties 
of atomic states in optical lattices. Such excitations play an important role 
in defining the boundary of the region where atoms can be localized. The presence
of phonons can provoke an instability of an insulating state, triggering atomic 
delocalization, often destroying the insulating state that would exist without 
these excitations. This delocalization effect can occur for both types of atomic 
interactions, long-range as well as local.

Throughout the paper, the system of units is employed, where the Planck and
Boltzmann constants equal to one $(\hbar = 1, k_B = 1)$.

\section{Main definitions and notations}

In this section, we give the main definitions and notations that will be used
in the following sections. 

We consider a fixed optical lattice described by the spatial points 
$\{ {\bf a}_i \}$ corresponding to $N_L$ lattice sites enumerated by the index 
$i = 1, 2, \ldots, N_L$. The elementary lattice cell is characterized by the set 
of vectors ${\bf a} = \{a_\alpha\}$, where the spatial components are enumerated 
by the index $\alpha = 1, 2, \ldots, d$. For the sake of generality, we consider 
a $d$-dimensional space, which will make it straightforward to analyze 
the particular cases of one- two- and three-dimensional lattices. 

The lattice contains $N$ atoms, whose ratio to the number of sites $N_L$ defines 
the filling factor
\be
\label{1}
 \nu \equiv \frac{N}{N_L} = \rho a^d \;  ,
\ee
in which $\rho$ is the average atomic density and $a$ is the mean interatomic
distance, given by the expressions
\be
\label{2}
 \rho \equiv \frac{N}{V} \; , \qquad
a  \equiv \left (\frac{V}{N_L} \right )^{1/d} \; ,
\ee
with $V$ being the system volume. The filling factor can be an arbitrary positive 
number. 

The optical lattice, formed by laser beams, is characterized by the lattice
potential
\be
\label{3}
V_L(\br) = \sum_{\al=1}^d V_\al \sin^2 \left ( k_0^\al r_\al \right )\; ,
\ee
where $k_0^\alpha = \pi / a_\alpha$. The recoil energy, playing the role of
a characteristic kinetic energy, is denoted as
\be
\label{4}
E_R = \frac{k_0^2}{2m} \qquad \left ( k_0^2 \equiv
\frac{1}{d} \sum_{\al=1}^d \frac{\pi^2}{a_\al^2} \right ) \; ,
\ee
with $m$ being atomic mass. The Schr\"{o}dinger lattice Hamiltonian is
\be
\label{5}
 H_L(\br) = -\; \frac{\nabla^2}{2m} + V_L(\br) \;  ,
\ee
with the lattice potential (3).

The extended Hubbard Hamiltonian is derived in the usual way. One starts
with the standard Hamiltonian of atoms with pair interactions $\Phi(\bf r)$,
expands the field operators over the Wannier functions $w({\bf r} - {\bf a}_i)$,
and restricts the consideration to the lowest energy band. For what follows,
we shall need the notations for the matrix elements over the Wannier functions:
the hopping term
\be
\label{6}
 J_{ij} \equiv -\int w^*(\br-\ba_i) H_L(\br) w(\br-\ba_j) \; d\br \;  ,
\ee
the intersite interaction
\be
\label{7}
  U_{ij} \equiv
\int  | w(\br-\ba_i)|^2 \Phi(\br-\br') | w(\br'-\ba_j) |^2 \; d\br d\br' \; ,
\ee
the momentum squared
\be
\label{8}
  \bp_j^2 \equiv \int w^*(\br-\ba_j) (-\nabla^2) w(\br-\ba_j) \; d\br \;  ,
\ee
and the average lattice parameter
\be
\label{9}
 V_L \equiv \int w^*(\br) V_L(\br) w(\br) \; d\br  \;  .
\ee
With these notations, one obtains the extended Hubbard Hamiltonian
\be
\label{10}
 \hat H = - \sum_{i\neq j} J_{ij} c_i^\dgr c_j + \sum_j \left (
\frac{\bp_j^2}{2m} + V_L \right )  c_j^\dgr c_j  +
\frac{U}{2} \sum_j c_j^\dgr c_j^\dgr c_j c_j +
\frac{1}{2} \sum_{i\neq j} U_{ij} c_i^\dgr c_j^\dgr c_j c_i \; ,
\ee
in which $U \equiv U_{jj}$. The constant term $V_L$ in Hamiltonian (10)
can be omitted. In the hopping term, it is customary to consider
only the nearest neighbors, denoting as $J$ the value of $J_{ij}$ related 
to these nearest neighbors. Neglecting the last term in Eq. (10), describing
intersite atomic interactions, would reduce the Hamiltonian to the standard 
Hubbard model. This omission can be motivated by the fact that usually the 
value of $U_{ij}$ for $i \neq j$ is smaller than the on-site interaction $U$. 
However, for the treatment of phonon excitations, the intersite interactions 
are crucial, even when they are small.

The atomic interactions, generally, contain two parts,
\be
\label{11}
\Phi(\br) = \Phi_{loc}(\br) + \Phi_{non}(\br) \;  ,
\ee
the local interactions, described by a delta function,
\be
\label{12}
 \Phi_{loc}(\br) = \Phi_d \dlt(\br) \;  ,
\ee
and non-local long-range interactions $\Phi_{non} (\bf r)$, such as dipolar
interactions \cite{Griesmaier_5,Baranov_6,Pupillo_7,Baranov_8}. The parameters
of the local and non-local interactions can be connected, but to a large extent,
the two types of the interactions can be treated as independent. Moreover, the
strengths of these interactions can be varied in a wide range. Respectively,
the interaction term (7) is a sum of two parts,
\be
\label{13}
 U_{ij} = U_{ij}^{loc} + U_{ij}^{non} \;  ,
\ee
corresponding to the local interactions,
\be
\label{14}
 U_{ij}^{loc} = \Phi_d  \int  | w(\br-\ba_i)|^2 | w(\br-\ba_j) |^2 \; d\br ,
\ee
and to non-local interactions,
\be
\label{15}
 U_{ij}^{non} =
\int | w(\br-\ba_i)|^2 \Phi_{non}(\br-\br') | w(\br'-\ba_j) |^2 \; d\br d\br' \;  .
\ee

The strength of the local interaction potential $\Phi_d$ depends on the system
dimensionality and setup geometry. Considering a system with atomic dynamics 
in $d$ dimensions, we can keep in mind the realistic situation, when $3-d$ 
directions are confined to the ground state of a harmonic oscillator, with a 
frequency $\omega_\perp$, to a size $l_\perp \equiv 1/\sqrt{m \omega_\perp}$. 
Then, we will have quasi-one-dimensional or quasi-two-dimensional systems
\cite{Petrov_9,Castin_10}.

For example, in three dimensions, the local-potential strength is
$$
\Phi_3 \equiv \Phi_0 = 4\pi \; \frac{a_s}{m} \;   ,
$$
where $a_s$ is the $s$-wave scattering length. For quasi-two dimensional
bosons \cite{Petrov_11}, one has
$$
 \Phi_2 \cong
\frac{\Phi_0}{\sqrt{2\pi}\; l_\perp - a_s \ln [ (2\pi)^{3/2}\rho l_\perp a_s] } \;  .
$$
And for quasi-one-dimensional bosons \cite{Olshanii_12}, one gets
$$
  \Phi_1 \cong \frac{\Phi_0}{ 2\pi l_\perp (l_\perp - 0.46 a_s)} \; .
$$
If the scattering length is much shorter than the length of the transverse
confinement, then the above equations reduce to the formula
$$
 \Phi_d =  \frac{\Phi_0}{(\sqrt{2\pi}\; l_\perp)^{3-d}} \qquad
 \left ( \frac{a_s}{l_\perp} \ll 1 \right ) \;  .
$$

All the cases considered above can be summarized in the form
\be
\label{16}
 \Phi_d =  \frac{\Phi_{eff}}{(\sqrt{2\pi}\; l_\perp)^{3-d}}  ,
\ee
in which the effective strength
\be
\label{17}
\Phi_{eff} = 4\pi \; \frac{a_{eff}}{m}
\ee
is expressed through the effective scattering length. The latter in the
quasi-one-dimensional case reads as
\be
\label{18}
 a_{eff} =  \frac{a_s}{1 - 0.46 a_s/l_\perp} \qquad ( d = 1 ) \;  ,
\ee
in the quasi-two-dimensional case, it is
\be
\label{19}
  a_{eff} =
\frac{a_s}{1 - (a_s/ \sqrt{2\pi}\;l_\perp) \ln [(2\pi)^{3/2}\rho l_\perp a_s] }
\qquad ( d = 2)  \;  ,
\ee
and in three dimensions, it reduces to
\be
\label{20}
 a_{eff} = a_s \qquad ( d = 3 ) \;  .
\ee

Above, we have listed the expressions that we will need in what follows. The 
derivation of these formulas can be found in the cited literature. Typical
expressions for the system parameters for an insulating state, such as the 
hopping term and intersite interactions, are given in the Appendix A.

\section{Vibrational collective excitations}

As is well known, collective excitations can essentially influence the system 
properties, defining the stability boundaries of different physical states
\cite{Birman_13}. Thus, the stability of an insulating state can depend on the 
existence of phonon excitations. These collective excitations can modify the 
properties of many-body systems, even when particle interactions are rather small. 
In particular, phonon excitations can destabilize the system leading to particle 
delocalization destroying an insulating state. As examples of systems where 
phonon excitations can strongly influence the region of stability, we can 
mention ferroelectrics \cite{Blinc_14,Yukalov_14} and atoms in double-well 
potentials \cite{Yukalov_15,Yukalov_16}. Here, we consider the role of phonons 
on the stability of insulating states in optical lattices. Here and in what 
follows, speaking about insulating states, we keep in mind the states typical 
of Mott insulators. I such states, atoms are well localized. But the notion 
of localization is more general and essentially depends on atomic interactions.
We employ the term "localization" in this general sense. 

Collective excitations are standardly introduced by considering small 
deviations from equilibrium values. Taking account of phonon excitations can 
be done in the traditional way \cite{Reissland_17,Bottger_18} by considering 
vibrating atoms characterized by vectors ${\bf r}_j$, oscillating around 
the related lattice sites ${\bf a}_j$, with introducing the deviations 
${\bf u}_j$ according to the rule
\be
\label{21}
 \br_j = \ba_j + \bu_j \;  ,
\ee
requiring the validity of the conditions for the averages
\be
\label{22}
 \ba_j \equiv \lgl \br_j \rgl \; , \qquad \lgl \bu_j \rgl = 0 \;  ,
\ee
where $\langle ...\rangle=Tr[... e^{-\beta H}]/Tr[e^{-\beta H}]$ is an average
with respect to the total Hamiltonian including all thermal and quantum
fluctuations. A vanishing average $\bu_j$ is based on the condition that
the system is in stable equilibrium, where $\ba_j$ is the definition of the
lattice vectors prescribed by the given equilibrium optical lattice. The 
deviation $\bu_j$ can become nonzero at the points of instability, such as 
the Peierls instability.

The quantities $U({\bf r}_{ij})$ and $J({\bf r}_{ij})$ depend on the difference
of the spatial variables
\be
\label{23}
\br_{ij} \equiv \br_i - \br_j = \ba_{ij} + \bu_{ij} \;   ,
\ee
where we use the notation
\be
\label{24}
 \bu_{ij} \equiv \bu_i - \bu_j \; , \qquad  \ba_{ij} = \ba_{i} -\ba_j \;   .
\ee

It is worth mentioning the main difference between the optical lattice, whose
periodicity and lattice vectors $\ba_i$ are strictly prescribed by the imposed
laser beams, and a self-organized crystal, whose lattice vectors $\ba_i$ are
defined self-consistently through the minimization of a thermodynamic potential. 

In a localized state, atomic deviations from the lattice sites are supposed
to be small, which justifies the expansion of the interaction potentials in
powers of the deviations. As usual, restricting such an expansion by the 
second order, we have
$$
U(\br_{ij}) \simeq U_{ij} + \sum_\al U_{ij}^\al u_{ij}^\al - \;
\frac{1}{2} \sum_{\al\bt} U_{ij}^{\al\bt} u_{ij}^\al u_{ij}^\bt \; ,
$$
\be
\label{25}
 J(\br_{ij}) \simeq J_{ij} + \sum_\al J_{ij}^\al u_{ij}^\al - \;
\frac{1}{2} \sum_{\al\bt} J_{ij}^{\al\bt} u_{ij}^\al u_{ij}^\bt \;  ,
\ee
where we use the notations
$$
U_{ij} \equiv U(\ba_{ij} ) \; , \qquad J_{ij} \equiv J(\ba_{ij} ) \;  ,
$$
$$
U_{ij}^\al \equiv \frac{\prt U_{ij}}{\prt a_i^\al} \; , \qquad
J_{ij}^\al \equiv \frac{\prt J_{ij}}{\prt a_i^\al} \; ,
\qquad
U_{ij}^{\al\bt} \equiv \frac{\prt^2 U_{ij}}{\prt a_i^\al\prt a_j^\bt} \; ,
 \qquad
J_{ij}^{\al\bt}  \equiv \frac{\prt^2 J_{ij}}{\prt a_i^\al \prt a_j^\bt} \; .
$$

As usual, to close the system of equations, it is necessary to decouple 
the high-order products of operators. These higher-order operator products, 
involving the variables of different nature, can be decoupled. Thus, we 
decouple the atomic and vibrational degrees of freedom involving the 
second-order vibrational variables:
$$
u_{ij}^\al u_{ij}^\bt c_i^\dgr c_j^\dgr c_j c_i =
\lgl u_{ij}^\al u_{ij}^\bt \rgl c_i^\dgr c_j^\dgr c_j c_i +
u_{ij}^\al u_{ij}^\bt \lgl  c_i^\dgr c_j^\dgr c_j c_i \rgl -
\lgl u_{ij}^\al u_{ij}^\bt \rgl \lgl  c_i^\dgr c_j^\dgr c_j c_i \rgl \; ,
$$
$$
u_{ij}^\al u_{ij}^\bt c_i^\dgr  c_j  =
\lgl u_{ij}^\al u_{ij}^\bt \rgl c_i^\dgr  c_j  +
u_{ij}^\al u_{ij}^\bt \lgl  c_i^\dgr  c_j \rgl -
\lgl u_{ij}^\al u_{ij}^\bt \rgl \lgl  c_i^\dgr  c_j \rgl \; ,
$$
\be
\label{26}
\bp^2_j  c_j^\dgr c_j = \lgl \bp^2_j \rgl c_j^\dgr c_j +
\bp^2_j  \lgl  c_j^\dgr c_j  \rgl  -
\lgl \bp^2_j \rgl \lgl  c_j^\dgr c_j  \rgl \; .
\ee
Such a decoupling is motivated by the different physical nature of the atomic
and deviation operators. Keeping in mind the lattice periodicity, the filling 
factor can be represented as 
\be
\label{27}
 \nu \equiv \frac{N}{N_L} = \frac{1}{N_L} \sum_j \lgl  c_j^\dgr c_j  \rgl
= \lgl  c_j^\dgr c_j  \rgl  .
\ee

Employing the above decouplings in the Hamiltonian, we meet the combination 
of terms, for which it is convenient to introduce the following notations.   
Thus, we define the effective hopping term
\be
\label{28}
 \widetilde J_{ij} \equiv   J_{ij} - \; \frac{1}{2}
\sum_{\al\bt} J_{ij}^{\al\bt} \lgl u_{ij}^\al u_{ij}^\bt \rgl
\ee
and the effective atomic interactions
\be
\label{29}
  \widetilde U_{ij} \equiv   U_{ij} - \; \frac{1}{2}
\sum_{\al\bt} U_{ij}^{\al\bt} \lgl u_{ij}^\al u_{ij}^\bt \rgl  ,
\ee
whose values are renormalized by the presence of atomic vibrations. These 
atomic vibrations are correlated with each other through the effective 
interaction matrix
\be
\label{30}
\Phi_{ij}^{\al\bt} \equiv U_{ij}^{\al\bt}  \lgl  c_i^\dgr c_j^\dgr c_j c_i \rgl
- 2 J_{ij}^{\al\bt} \lgl c_i^\dgr c_j \rgl  \;  .
\ee
Atoms produce the effective deformation force
\be
\label{31}
 F_{ij}^{\al} \equiv - U_{ij}^\al c_i^\dgr c_j^\dgr c_j c_i
+  2 J_{ij}^\al c_i^\dgr c_j \;  ,
\ee
caused by atom-vibration correlations. It is also important to notice that 
for a function $f({\bf a}_{ij})$, depending on the difference ${\bf a}_{ij}$, 
the following properties are valid:
\be
\label{32}
\sum_j \frac{\prt f(\ba_{ij})}{\prt a_i^\al} =
\frac{\prt}{\prt a_i^\al} \sum_j f(\ba_{ij}) = 0 \; ,
\qquad
  \sum_j \frac{\prt^2 f(\ba_{ij})}{\prt a_i^\al \prt a_j^\bt} =
-\; \frac{\prt}{\prt a_i^\al}
\sum_j \frac{\prt f(\ba_{ij})}{\prt a_{ij}^\bt} = 0 \;  .
\ee
These properties will be used in the final presentation of the system Hamiltonian. 

Accomplishing the described procedure for Hamiltonian (10), we obtain
\be
\label{33}
 \hat H = E_N + \hat H_{at} + \hat H_{vib} + \hat H_{int} \;  .
\ee
Here the first term is the non-operator quantity
\be
\label{34}
E_N = \frac{1}{4} \sum_{i\neq j}
\sum_{\al\bt} \Phi_{ij}^{\al\bt} \lgl u_{ij}^\al u_{ij}^\bt \rgl -
\nu \sum_j \; \left \lgl \frac{\bp_j^2}{2m} \right \rgl \; .
\ee
Atoms are described by the renormalized Hamiltonian
\be
\label{35}
 \hat H_{at} = - \sum_{i\neq j} \widetilde J_{ij}  c_i^\dgr c_j +
  \frac{U}{2} \sum_j  c_j^\dgr  c_j^\dgr c_j c_j +
\frac{1}{2} \sum_{i\neq j} \widetilde U_{ij}  c_i^\dgr c_j c_j c_i +
\sum_j \; \left \lgl \frac{\bp_j^2}{2m} \right \rgl c_j^\dgr  c_j \; .
\ee
Collective vibrational degrees of freedom are characterized by the 
Hamiltonian
\be
\label{36}
\hat H_{vib} =  \nu \sum_j  \frac{\bp_j^2}{2m} - \frac{1}{4}
\sum_{i\neq j} \sum_{\al\bt} \Phi_{ij}^{\al\bt} u_{ij}^\al u_{ij}^\bt \;  .
\ee
And the last term $\hat H_{int}$ corresponds to local deformations caused 
by the correlations between atomic and vibrational degrees of freedom,
\be
\label{37}
  \hat H_{int} = -\;
\frac{1}{2} \sum_{i\neq j} \sum_\al F_{ij}^\al u_{ij}^\al \;  .
\ee
Passing from the relative deviations ${\bf u}_{ij}$ to the single-site 
deviations ${\bf u}_j$, and using the above properties, the vibrational Hamiltonian 
part can be represented as
\be
\label{38}
\hat H_{vib} = \nu \sum_j  \frac{\bp_j^2}{2m} + \frac{1}{2}
\sum_{i\neq j} \sum_{\al\bt} \Phi_{ij}^{\al\bt} u_i^\al u_j^\bt \;  .
\ee
The effective deformation force (31) enjoys the property
\be
\label{39}
 F_{ji}^\al = - F_{ij}^\al  \qquad (i \neq j) \; .
\ee
Therefore the deformation term, caused by the correlations between atomic 
and vibrational degrees of freedom, can be rewritten as
\be
\label{40}
  \hat H_{int} = \sum_{i\neq j} \sum_\al F_{ij}^\al u_j^\al \;  .
\ee
Thus, all terms of the Hamiltonian $\hat{H}$ are defined.

Let us stress that the vibrational collective excitations appear only when
there exist intersite atomic interactions correlating atoms. In the presence
of these interactions, atoms move in an effective potential composed of 
an optical lattice and a self-organized field formed by intersite interactions.
The optical lattice does not correlate atoms, prescribing only their individual
properties. While intersite atomic interactions do collectivize the atoms, 
whose collective vibrations play the role of collective phonon excitations.

\section{Quantization of phonon variables}

Quantized phonon variables are introduced so that to diagonalize the part 
of the Hamiltonian containing atomic deviations. In our case, the difference 
with the standard introduction of phonon operators is due to the existence of
the linear in the deviation term (40). Dealing with such linear terms
requires to slightly modify the corresponding canonical transformation
\cite{Yukalov_19}. In that case, the phonon operators are introduced by means 
of the nonuniform transformation
$$
\bu_j = \vec{\Dlt}_j +
\frac{1}{\sqrt{2N}} \sum_{ks} \sqrt{\frac{\nu}{m\om_{ks}} } \; \bfe_{ks}
\left ( b_{ks} + b^\dgr_{-ks} \right ) e^{i\bk \cdot \ba_j} \; ,
$$
\be
\label{41}
\bp_j = -\;
\frac{i}{\sqrt{2N}} \sum_{ks} \sqrt{\frac{m\om_{ks}}{\nu } } \; \bfe_{ks}
\left ( b_{ks} - b^\dgr_{-ks} \right ) e^{i\bk \cdot \ba_j} \;    ,
\ee
in which ${\bf e}_{ks}$ are the polarization vectors, with $s$ being the
polarization index. The phonon frequencies are given by the eigenproblem
\be
\label{42}
 \frac{\nu}{m} \sum_{j(\neq i)} \sum_{\bt}
\Phi_{ij}^{\al\bt} e^{i\bk \cdot \ba_{ij}} e_{ks}^\bt = \om^2_{ks} e^\al_{ks} \; ,
\ee
with the effective interaction matrix (30). Diagonalizing the phonon part
of the Hamiltonian yields
\be
\label{43}
 \Dlt_i^\al = \sum_{j(\neq i)} \sum_{\bt} \gm_{ij}^{\al\bt} F_j^\bt \; ,
\ee
where we use the notation
\be
\label{44}
\gm_{ij}^{\al\bt} \equiv  \frac{\nu}{N}
\sum_{ks} \frac{e_{ks}^\al e_{ks}^\bt}{m \om^2_{ks}}\;
 e^{i\bk \cdot \ba_{ij}}
\ee
and where the effective deformation force, acting on an atom, is
\be
\label{45}
  F_i^\al = \sum_{j(\neq i)} F_{ij}^\al \; .
\ee
The presence of the term $\vec{\Dlt}_j$ in the canonical transformation (41)
distinguishes the latter from the standard canonical transformation in the 
quantization of phonon variables.   

It is easy to see that variables (41) satisfy the usual commutation relations
$$
 [ u_i^\al , \; p_j^\bt   ] =  i \dlt_{ij} \dlt_{\al\bt} \;   .
$$
Conditions (22) are valid, since
\be
\label{46}
 \lgl \vec{\Dlt}_i \rgl = 0 \;  .
\ee

Then Hamiltonian (33) results in the sum
\be
\label{47}
  \hat H = E_N + \hat H_{at} + \hat H_{ph} + \hat H_{ind} \; .
\ee
Here the first term is the same as in Eq. (34). The atomic Hamiltonian 
is given by Eq. (35). The phonon Hamiltonian is diagonal,
\be
\label{48}
 \hat H_{ph} =
\sum_{ks} \om_{ks} \left ( b_{ks}^\dgr b_{ks} + \frac{1}{2} \right ) \;  .
\ee
And the last term is the Hamiltonian of effective multi-atomic interactions,
induced by atomic vibrations,
\be
\label{49}
 \hat H_{ind} = 
\sum_{i\neq j} \sum_{\al\bt} F_i^\al \gm^{\al\bt}_{ij} F_j^\bt \; .
\ee

Using properties (32), it is straightforward to check that the average 
force (45) is zero, that is, $\langle F_j^\alpha \rangle = 0$. Hence, the 
induced term (49), in the mean-field approximation is zero, 
$\langle {\hat H}_{ind} \rangle = 0$, since
$$
\lgl F_i^\al  F_j^\bt \rgl \cong \lgl F_i^\al  \rgl \lgl F_j^\bt \rgl = 0 \; .
$$
Moreover, the induced term (49) is much smaller compared to the 
atomic term (35). This implies that the induced term (49) does not 
influence much the properties of atoms.

The properties of phonons depend on their frequency, for which we have
the equation
\be
\label{50}
 \om^2_{ks} = \frac{\nu}{m} \sum_{j(\neq i)} \sum_{\al\bt}
\Phi_{ij}^{\al\bt} e_{ks}^\al  e_{ks}^\bt e^{i\bk \cdot \ba_{ij}}\; .
\ee
The deviation correlation function becomes
\be
\label{51}
 \lgl u_i^\al u_j^\bt \rgl = \dlt_{ij} \frac{\nu}{2N} \sum_{ks}
\frac{e_{ks}^\al e_{ks}^\bt}{m\om_{ks}}
 \coth \left ( \frac{\om_{ks}}{2T} \right ) \;  .
\ee
And the average kinetic energy per atom is
\be
\label{52}
 \left \lgl \frac{\bp_j^2}{2m} \right \rgl =
\frac{1}{4\nu N} \sum_{ks} \om_{ks}
 \coth \left ( \frac{\om_{ks}}{2T} \right ) \;  .
\ee

An important quantity, characterizing the width of atomic vibrations,  
is the mean-square deviation $r_0$ defined by the equation
\be
\label{53}
 r_0^2 \equiv \sum_{\al=1}^d \lgl u_i^\al u_i^\al \rgl \;  .
\ee
This quantity should not be confused with $l_0$ that defines the width 
of a wave packet at a lattice site. While $r_0$ is the mean deviation of 
this wave packet oscillating around the lattice site.

In the case of a $d$-dimensional {\it cubic} lattice, the frequency is the 
same for all polarizations, which can be described by the relation
\be
\label{54}
  \om_k^2 = \frac{1}{d} \sum_{s=1}^d \om_{ks}^2 \; .
\ee
The latter results in the equation for the phonon frequency
\be
\label{55}
 \om_k^2 = -\; \frac{\nu}{m} \sum_{j(\neq i)} D_{ij} e^{i\bk \cdot \ba_{ij}}  ,
\ee
with the dynamical matrix
\be
\label{56}
 D_{ij} \equiv - \; \frac{1}{d} \sum_{\al=1}^d \Phi_{ij}^{\al\al} =
\frac{1}{d}
\sum_{\al=1}^d \frac{\prt^2\Phi_{ij}}{\prt a_i^\al \prt a_i^\al} \;  .
\ee

For a cubic lattice, the mean-square deviation (53) reads as
\be
\label{57}
  r_0^2 = \frac{\nu d}{2m\rho} \int_{\cal{B}} \frac{1}{\om_k}
\coth \left ( \frac{\om_k}{2T} \right ) \frac{d\bk}{(2\pi)^d} \;   ,
\ee
with the integration over the Brillouin zone. Taking into account only the
nearest-neighbor interactions leads to the effective phonon dispersion
\be
\label{58}
 \om_k^2 =
\frac{4\nu}{m} \; D_0 \sum_\al \sin^2 \left ( \frac{k_\al a}{2} \right ) \;  ,
\ee
where $a_\alpha = a$ and $D_0$ is $D_{ij}$ for the nearest neighbors. In the
long-wave limit, Eq. (58) reduces to the acoustic spectrum
\be
\label{59}
  \om_k \simeq c_0 k \qquad
\left ( k^2 \equiv \sum_{\al=1}^d k_\al^2 \; \ra \; 0 \right ) \; ,
\ee
with the sound velocity
\be
\label{60}
c_0 = \sqrt{ \frac{\nu}{m} \; D_0 a^2 } \;   .
\ee

The value of the mean-square deviation (57) describes the properties of the 
localized state and defines the region, where it can exist. As is clear, to 
be treated as localized, the state has to enjoy a mean-square deviation that
would be much smaller than the mean interatomic distance. This is the essence
of the Lindemann criterion of stability that will be considered in the next
section.

\section{Possibility of phonon instability}
\label{sect:inst}

One of the most important characteristics of a localized solid-like state
is the mean-square deviation (53) or (57). A localized state can exist only 
when this deviation $r_0$ is much smaller than the distance $a$ between the 
nearest neighbors. This statement is the well known {\it Lindemann criterion 
of lattice stability} (see, e.g., \cite{Reissland_17,Bottger_18,Ziman_20}). 
According to this criterion, the majority of solids become unstable and melt 
when the Lindemann ratio $r_0/a$ surpasses $0.2$. This criterion is valid for 
anharmonic crystals as well \cite{Leibfried_21}. Even for such a strongly 
anharmonic quantum crystal as $^3$He the Lindemann ratio is $0.3$, which is 
measured experimentally \cite{Guyer_22}. In the weakest form, the Lindemann 
criterion \cite{Lindemann_23} states that for the stability of a localized 
solid-like system it is necessary that
\be
\label{61}
 \frac{r_0}{a} < 1 \;  .
\ee
The meaning of the Lindemann criterion is evident: If the mean-square 
deviations of neighboring atoms would be comparable to their mean interatomic 
distance, the system could not be considered as localized. 

In the long-wave limit, the phonon spectrum is acoustic, as is shown in Eq. (59).
This tells us that, calculating the mean-square deviation (57), the limit of 
small wave vectors can produce infrared divergence, depending on the system 
dimensionality and temperature. In order to study when and how this happens,
we can employ the standard procedure of limiting integral (57) from below by 
introducing the minimal wave vector $k_{min}$ that is assumed to tend to zero.
Equivalently, it is possible to define the minimal wave vector as 
$k_{min} = \pi/L$, where $L = a N_L^{1/d}$ is the length of the lattice. As is
clear, for a large lattice, with the number of lattice sites $N_L \ra \infty$,
the minimal wave vector tends to zero.  

Considering integral (57) at finite temperatures $T > 0$ for low dimensionality 
$d < 2$ shows that the mean-square deviation diverges as
$$
 r_0^2 \simeq \frac{T N_L^{2/d-1}d}{2^d \pi^2 (2-d)\nu D_0} \qquad
( d < 2 , \; T > 0 ) \;  ,
$$
where $D_0$ is the dynamical matrix, as in Eq. (58), and $N_L \ra \infty$. 
In particular, for one-dimensional space ($d = 1)$, the divergence is linear
in the number of sites $N_L$, 
\be
\label{62}
  r_0^2 \simeq \frac{T N_L}{2\pi^2 \nu D_0} \qquad ( d = 1 , \; T > 0 ) \;  .
\ee
And for two-dimensional space ($d = 2$), the divergence is logarithmic in $N_L$,
\be
\label{63}
 r_0^2 \simeq \frac{T \ln N_L}{(2\pi)^2 \nu D_0} \qquad ( d = 2 , \; T > 0 ) \;  .
\ee
This means that at finite temperatures the localized state is unstable in
dimensions $d = 1$ and $d = 2$ for asymptotically large lattices, where 
$N_L \ra \infty$. Although the lattice could exist for such $N_L$ that would 
be large, but finite, at the same time satisfying the Lindemann criterion (61).  

Let us notice that the dynamical matrix $D_0$, of course, depends on the 
parameters of the optical lattice. This will be shown below by explicit 
equations. However the infrared divergence, considered above, is the feature 
typical of low-dimensional systems. Recall that, according to the Mermin-Wagner 
theorem \cite{Mermin_23,Hohenberg_23,Coleman_23} continuous symmetry at finite 
temperature cannot be broken in the spaces of dimensionality lower than three 
($d \leq 2$), irrespectively of the strength of their interactions, provided 
that the latter are of short-range type. Such a low-dimensional infrared 
instability is a purely dimensional effect. However, this low-dimensional 
effect is not of great danger for real world that is three-dimensional. Actually,
dealing with optical lattices, one always deals with three-dimensional systems,
which can be reduced to {\it quasi-one-dimensional} or {\it quasi-two-dimensional}
by integrating out some degrees of freedom.     

It is generally accepted that for solid states the Debye approximation gives
a quite accurate description
\cite{Reissland_17,Bottger_18,Ziman_20,Leibfried_21,Guyer_22}. In this
approximation, the integration over the Brillouin zone is replaced by the
integration over the Debye sphere,
\be
\label{64}
 \int_{\cal{B}}  \frac{d\bk}{(2\pi)^d} \; \ra \;
\frac{2}{(4\pi)^{d/2}\Gm(d/2)} \int_0^{k_D} k^{d-1} \; dk \;  ,
\ee
limited by the Debye radius $k_D$ that is defined by the normalization
condition
\be
\label{65}
   \int_{\cal{B}}  \frac{d\bk}{(2\pi)^d} = \frac{N_L}{V} = \frac{\rho}{\nu} \; ,
\ee
where $\rho$ is the average density
\be
\label{66}
\rho \equiv \frac{N}{V} = \frac{\nu}{a^d} \;   .
\ee
This gives the Debye radius
\be
\label{67}
 k_D = \frac{\sqrt{4\pi}}{a} \left [
\frac{d}{2} \; \Gm\left ( \frac{d}{2} \right ) \right ]^{1/d} \;  .
\ee
The spectrum is taken to be isotropic, with the frequency
\be
\label{68}
 \om_k = c_0 k \qquad ( 0 \leq k \leq k_D ) \;  ,
\ee
whose upper limit defines the Debye temperature
\be
\label{69}
 T_D \equiv c_0 k_D \;  .
\ee
With the notation for the Debye radius (67), the replacement (64) takes 
the form
\be
\label{70}
 \int_{\cal{B}}  \frac{d\bk}{(2\pi)^d} \; \ra \;
\frac{d}{(k_D a)^d} \int_0^{k_D} k^{d-1}\; dk \;   .
\ee
Below, we study in more detail the Lindemann criterion, employing the 
Debye approximation. 

At zero temperature, the mean-square deviation becomes
\be
\label{71}
 r_0^2 = \frac{d^2}{2(d-1)mT_D} \qquad ( T = 0 ) \;  .
\ee
The Lindemann criterion of stability (61) yields
\be
\label{72}
   T_D > \frac{d^2}{2(d-1)m a^2} \qquad ( T = 0 ) \; .
\ee
The meaning of the latter inequality is very transparent: a localized state
can be formed only when the effective potential energy is larger than the
kinetic energy of atoms. As Eq. (72) demonstrates, no localized state 
can exist at zero temperature for $d = 1$.

At high temperatures, the mean-square deviation is given by the expression
\be
\label{73}
  r_0^2 \simeq \frac{Td^2}{(d-2)mT_D^2} \qquad ( T \gg T_D  ) \; .
\ee
Then the Lindemann stability criterion gives
\be
\label{74}
   T_D > \sqrt{\frac{Td^2}{(d-2)m a^2} } \qquad ( T \gg T_D  ) \; .
\ee
This tells us that, at such temperatures, there can be no localized state
for $d = 2$. At these temperatures, only a three-dimensional localized state
can exist, provided that
\be
\label{75}
  T_D > \sqrt{\frac{9T}{m a^2} } \qquad ( T > T_D , \; d = 3 ) \;  .
\ee

To simplify the consideration, for a well localized insulating state, one can 
use the averages $\lgl c_i^\dgr c_j \rgl = \dlt_{ij} \nu$ and
\be
\label{76}
\lgl c_i^\dgr c_j^\dgr  c_j   c_i \rgl =
\lgl c_i^\dgr c_i \rgl \lgl c_j^\dgr c_j \rgl = \nu^2 \qquad (i \neq j) \;  .
\ee
As a result, the interaction matrix (30) reduces to
\be
\label{77}
 \Phi_{ij}^{\al\bt} = \nu^2 U_{ij}^{\al\bt} \;  .
\ee

An interesting question is how atomic vibrations influence a localized state
due to local interactions. For such a case, considering a cubic lattice, 
taking account of only the nearest neighbors, and involving the formulas 
from the Appendix A, we get
\be
\label{78}
  U_{ij} = U \exp \left ( - \; \frac{a^2 d}{2l_0^2} \right ) \;  ,
\ee
where the equality $a_{ij}^2 = a^2 d$ is used and $U = U_{loc}$ is defined
in the Appendix A. The dynamical matrix (56) becomes
\be
\label{79}
 D_0 = \left ( \frac{\nu a }{l_0^2} \right )^2
U \exp \left ( - \; \frac{a^2 d}{2l_0^2} \right ) \; .
\ee
The sound velocity (60) reads as
\be
\label{80}
 c_0 = \frac{\nu a^2}{l_0^2} \; \sqrt{ \frac{\nu}{m} \; U }
\exp \left ( - \; \frac{a^2 d}{4l_0^2} \right ) \;  ,
\ee
which can also be represented in the form
\be
\label{81}
  c_0 = \frac{4\nu Ja^2}{(\pi^2-4)V_0l_0^2 d} \;
\sqrt{ \frac{\nu}{m} \; U } \; ,
\ee
through the parameters of the optical lattice.

For a cubic lattice at zero temperature, the Lindemann criterion of 
stability (61) can be written as
\be
\label{82}
\frac{ 16(2\pi)^{1/4}\nu^{3/2}J }{(\pi^2-4)V_0 d} \;
\left [ \frac{d}{2} \; \Gm \left ( \frac{d}{2} \right ) \right ]^{1/d}
\left ( \frac{a}{l_0} \right )^3 \;
\sqrt{ \frac{ a_{eff} }{ l_\perp^{3-d} \; l_0^{d-2} }  } \; > \;
\frac{d^2}{d-1}  \; .
\ee
Consequently, the system stability essentially depends on the lattice 
parameters, space dimensionality, as well as on atomic interactions.   

At zero temperature, there can arise only two- and three-dimensional localized
states. Considering the two-dimensional case $(d = 2)$, we have
$$
\Phi_2 = \frac{\Phi_{eff}}{\sqrt{2\pi}\; l_\perp} \; , \qquad
k_D = \frac{\sqrt{4\pi}}{a} \; ,
$$
\be
\label{83}
U = \frac{\Phi_{eff}}{(2\pi)^{3/2} l_\perp l_0^2 } \qquad (d = 2 ) \;   .
\ee
The Debye temperature (69) becomes
\be
\label{84}
 T_D = 2(2\pi)^{1/4}\;
 \frac{\nu a}{ml_0^3} \; \sqrt{\frac{\nu a_{eff}}{l_\perp} } \;
\exp \left ( - \; \frac{a^2}{2l_0^2} \right ) \;  .
\ee
The Lindemann criterion (72) is equivalent to the inequality
$$
 m a^2 T_D > 2 \;  ,
$$
which gives
\be
\label{85}
 \nu^{3/2} \left ( \frac{a}{l_0} \right )^3 \sqrt{\frac{a_{eff}}{l_\perp} } \;
\exp \left ( - \; \frac{a^2}{2l_0^2} \right )  > 0.632 \;   .
\ee

Comparing the ratio $U_{loc}/J$, given in the Appendix A, with 
criterion (85), we see that the atomic state can be insulating, when 
no phonon degrees of freedom are taken into account and $U$ is much larger 
than $J$. But as soon as phonon excitations are included, the stability 
condition (85), for the same parameters, may be not valid, which means 
that atoms delocalize. 

For the three-dimensional case $(d = 3)$, we get
$$
\Phi_3 = \Phi_0 \; , \qquad k_D = \frac{(6\pi^2)^{1/3}}{a} \; ,
$$
\be
\label{86}
 U = \frac{\Phi_0}{(2\pi)^{3/2} l_0^3} \qquad ( d = 3) \;  .
\ee
The Debye temperature (69) becomes
\be
\label{87}
  T_D = 3.482 \; \frac{\nu a}{ml_0^3} \; \sqrt{\frac{\nu a_s}{l_0} }
\exp \left ( - \; \frac{3a^2}{4l_0^2} \right ) \;  .
\ee
At zero temperature, the Lindemann criterion (72) takes the form
$$
ma^2 T_D > \frac{9}{4} \;   ,
$$
which yields
\be
\label{88}
\nu^{3/2}  \left ( \frac{a}{l_0} \right )^3 \sqrt{\frac{a_s}{l_0} } \;
 \exp \left ( - \; \frac{3a^2}{4l_0^2} \right ) > 0.648 \;  .
\ee
Again, the system can be insulating without phonons, but becoming delocalized
in the presence of the latter.

Using the formulas of the Appendix A, the localization conditions (85) and (88) 
can be represented in another form by taking into account the expressions for 
the hopping rate
$$
J = 2.935 \; V_0 \exp  \left ( - \; \frac{a^2}{2l_0^2} \right ) \qquad
(d = 2 ) \; ,
$$
$$
J = 4.402 \; V_0 \exp  \left ( - \; \frac{3a^2}{4l_0^2} \right ) \qquad
(d = 3 ) \; .
$$
Then Eq. (85) yields
\be
\label{89}
  \frac{a_{eff}}{l_\perp} > \frac{3.441}{\nu^3}
\left ( \frac{V_0}{J} \right )^2  \left ( \frac{l_0}{a} \right )^6
\qquad ( d = 2 ) \;  ,
\ee
while criterion (88) gives
\be
\label{90}
\frac{a_s}{l_0} > \frac{8.137}{\nu^3}
\left ( \frac{V_0}{J} \right )^2  \left ( \frac{l_0}{a} \right )^6
\qquad ( d = 3 ) \;   .
\ee
This shows that the Lindemann criterion of stability requires that atomic
interactions, hence the scattering length, be sufficiently large. If these
conditions are not valid, the insulating state can be destroyed by phonon 
vibrations.

In order to better understand the physics of the phonon instability, we need 
to analyze how the occurrence of atomic vibrations influences the values
of the hopping parameter and interaction matrix.

\section{Renormalization of atomic parameters}

Phonon excitations renormalize atomic parameters according to Eqs. (28)
and (29). The renormalized quantities are shifted, because of the phonon
existence, resulting in
\be
\label{91}
 \widetilde J_{ij} = J_{ij} + \Dlt J_{ij} \; , \qquad
  \widetilde U_{ij} = U_{ij} + \Dlt U_{ij} \;  ,
\ee
where, taking into account that
$$
\lgl u_{ij}^\al u_{ij}^\bt  \rgl =
2(1-\dlt_{ij} )  \lgl u_{i}^\al u_{j}^\bt  \rgl \;  ,
$$
we have
\be
\label{92}
 \Dlt J_{ij} = - \sum_{\al\bt} J_{ij}^{\al\bt} \lgl u_{j}^\al u_{j}^\bt  \rgl \; ,
\qquad
\Dlt U_{ij} = - \sum_{\al\bt} U_{ij}^{\al\bt} \lgl u_{j}^\al u_{j}^\bt  \rgl \;  .
\ee

In the Debye approximation, we find
\be
\label{93}
  \lgl u_{j}^\al u_{j}^\bt  \rgl = \dlt_{\al\bt} \; \frac{r_0^2}{d} \;  ,
\ee
which gives
\be
\label{94}
 \Dlt J_{ij} = -\; \frac{r_0^2}{d} \; \sum_\al J_{ij}^{\al\al} \; ,
\qquad
 \Dlt U_{ij} = -\; \frac{r_0^2}{d} \; \sum_\al U_{ij}^{\al\al} \; .
\ee

In the tight-binding approximation, where $l_0 \ll a$, the hopping parameter
shift is
\be
\label{95}
 \Dlt J_{ij} = \frac{a^2r_0^2}{4l_0^4} \; J_{ij} \;  ,
\ee
which shows that the hopping parameter increases due to collective phonon 
excitations. At zero temperature, the relative shift is 
\be
\label{96}
 \frac{\Dlt J_{ij}}{J_{ij}} =  \frac{a^2 d^2}{8(d-1) ml_0^4 T_D} \;  .
\ee
For a two-dimensional lattice $(d = 2)$, this gives
\be
\label{97}
 \frac{\Dlt J_{ij}}{J_{ij}} = 0.079 \; \frac{a}{\nu l_0} \;
\sqrt{ \frac{l_\perp}{\nu a_{eff}} } \;
\exp \left ( \frac{a^2}{2l_0^2} \right )
= 0.074 \; \frac{aV_0}{\nu^{3/2}l_0J}\;
\sqrt{\frac{l_\perp}{a_{eff}} }  \;  ,
\ee
and for a three-dimensional lattice $(d = 3)$, the relative shift is
\be
\label{98}
  \frac{\Dlt J_{ij}}{J_{ij}} = 0.054 \; \frac{a}{\nu l_0} \;
\sqrt{\frac{l_0}{\nu a_s} }\; \exp \left ( \frac{3a^2}{4l_0^2} \right )
= 0.237 \; \frac{aV_0}{\nu^{3/2}l_0J}\; \sqrt{\frac{l_0}{a_s} }  \;  .
\ee

The increase of the hopping parameter can be rather noticeable. Thus, for
$r_0/l_0 \sim 0.3$ and $l_0/a \sim 0.1$, shift (95) is of the order 
of $J_{ij}$. 

The shift of the interaction matrix, with the local interaction potential,
reads as
\be
\label{99}
 \Dlt U_{ij} = \frac{a^2r_0^2}{l_0^4} \; U_{ij} \;  ,
\ee
which means that interaction matrix increases. At zero temperature, this
yields
\be
\label{100}
\frac{\Dlt U_{ij}}{U_{ij}} =  \frac{a^2 d^2}{2(d-1) ml_0^4 T_D} \;   .
\ee

The increase of the effective interaction matrix $U_{ij}$, caused by collective
phonon excitations, can be sufficiently large, even in the case of the local 
atomic interactions. For instance, if $r_0/l_0 \sim 0.3$ and $l_0/a \sim 0.1$, 
then the shift (99) is of the order of $U_{ij}$. Thus, the phonon vibrations 
can essentially renormalize the atomic Hamiltonian parameters. A simple example 
of temperature dependence is given in Appendix B. 

In this way, collective phonon excitations lead to the increase of the hopping
parameter and of the interaction matrix corresponding to the interactions of 
atoms in different lattice sites. However, notice that the on-site atomic 
interaction parameter $U$ in Hamiltonian (35) remains unchanged.

\section{Physics of phonon instability}

Now it is straightforward to understand why collective phonon excitations
can lead to the instability of an insulating state. To this end, keeping in mind 
low temperatures, let us compare the energy of the system described by the 
Hamiltonian without phonon degrees of freedom with the energy of the system 
including phonon excitations. These energies are given by the average values 
of the related Hamiltonians. The energy of the system with phonons is defined 
as the average $\tilde{E} \equiv \langle \hat{H} \rangle$ of Hamiltonian (33). 
While the energy $E$ of the system without phonons can be defined as the average 
of Hamiltonian (10). So, we need to consider the difference 
$\Delta \equiv \tilde{E} - E$.    

As is explained above, when analyzing the interaction Hamiltonian (40), we find 
that its average value is small, $\langle \hat{H}_{int} \rangle \approx 0$. The 
Appendix A shows that the intersite term $U_{ij}$, with $i \neq j$ is much smaller 
than the on-site term $U$. Thus, we obtain the energy difference 
$$
\Dlt E \approx - \sum_{i\neq j} \Dlt J_{ij} \lgl c_i^\dgr c_j \rgl \;   .
$$       
In the previous section, it is shown that the hopping parameter shift is positive.
Thus, the existence of phonons increases the hopping parameter and, consequently,
decreases the system energy. Since the system with phonons prefers a smaller energy,
it is more stable than the system without these excitations. 

A localized state is metastable, since in order to destroy it, atoms have to 
penetrate through the barrier $V_0$ created by the optical lattice. Such a 
penetration requires some time that characterizes the lifetime of the metastable 
state. This lifetime can be estimated 
\cite{Sewell_23,Horsthemke_23,Gardiner_23,Hanggi_23,Talkner_23} as 
$$
 t_{met} = \tau_0 \exp \left ( \frac{V_B}{\ep_N} \right ) \;  ,
$$
where $V_B$ is the barrier height, $\varepsilon_N$ is the characteristic energy
of noise, which is defined by the characteristic kinetic energy, and $\tau_0$
is the period of atomic oscillations at the bottom of the well. For classical 
systems the characteristic noise energy $\varepsilon_N$ coincides with 
temperature $T$, which results in the Arrhenius formula. For quantum systems,
the characteristic kinetic energy is defined by the energy that, in our case, 
is the recoil energy $E_R$. And the barrier height for an optical lattice is $V_0$. 
With the oscillation period $\tau = 2 \pi / \omega_0$, we have the lifetime
$$
t_{met} = \frac{2\pi}{\om_0} \; \exp \left ( \frac{V_0}{E_R} \right ) \;  .
$$
    
As an illustration of the above consideration, let us study some experiments, 
where the insulating state in an optical lattice has been observed. For example, 
let us consider the experiment of Greiner et al. \cite{Greiner_23}, where a
three-dimensional cubic lattice was created, loaded with the atoms of $^{87}$Rb, 
with the filling number close to one. Varying the optical-lattice parameters, 
both the localized insulating and delocalized superfluid states were realized. 
The lattice was formed by laser beams of the wavelength 
$\lambda = 0.852 \times 10^{-4}$ cm, which makes the lattice parameter
$a = \lambda/2 = 0.426 \times 10^{-4}$ cm. With the mass 
$m = 1.443 \times 10^{-22}$ g, the recoil energy is
$E_R = 2.093 \times 10^{-23}$ erg. At the lattice depth
$V_0 = 13 E_R = 2.721 \times 10^{-22}$ erg, atoms are localized in a Mott state.
Under these parameters, the effective frequency
$\omega_0 = 1.509 \times 10^{-22}$ erg, which gives the wave packet width
$l_0 = 0.715 \times 10^{-5}$ cm. Since $l_0/a = 0.168$, the tight-binding
approximation is applicable. With the scattering length
$a_s = 0.545 \times 10^{-6}$ cm, we have the ratio $a_s/l_0 = 0.076$.

At the same time, for $V_0 = 13 E_R$ numerical calculations \cite{Greiner_23}
give $U = 0.35 E_R$ and $J = U/36 = 0.97 \times 10^{-2} E_R$. Then the Lindemann
criterion (90) requires that $a_s/l_0$ be larger than 330. According to this 
criterion, the localized insulating state is unstable if the phonon degrees 
of freedom are taken into account. Therefore, the experimentally observed
localized states of atoms interacting through local delta-function forces, are
not absolutely stable equilibrium states, but rather are metastable states.

The oscillation time of an atom in a well is 
$\tau = 2 \pi/ \omega_0 = 0.44 \times 10^{-3}$ s. Therefore, the lifetime of
a metastable insulating state, under $V_0 = 13 E_R$ is $t_{met} \approx 200$ s. 
This is a quite long time, allowing for its easy observation. Moreover, this 
time is longer than the typical lifetime of atoms in a trap, which is of order 
of seconds or tens of seconds \cite{Ketterle,Letokhov}. Therefore, it is very 
feasible to create long-lived metastable insulating states of trapped atoms 
in optical lattices.

\section{Atoms with dipole interactions}

In the case of non-local long-range atomic interactions $\Phi_{non} (\bf r)$,
such as dipolar interactions, atoms in an optical lattice, where $l_0 \ll a$,
can be described \cite{Griesmaier_5,Baranov_6,Pupillo_7,Baranov_8} by the
extended Hubbard model with the effective interactions
\be
\label{101}
 U_{ij} = \frac{C_D}{a_{ij}^3} \qquad ( i \neq j ) \;  ,
\ee
in which $C_D \propto \mu_0^2$, with $\mu_0$ being the magnetic (or electric)
atomic dipole. Since dipole interactions are of hard-core type, each lattice
site can host only a single atom, that is, the filling factor is one, $\nu = 1$.
For the interactions of form (101), one has
$$
  U_{ij}^\al = - 3 C_d \; \frac{a_{ij}^\al}{a_{ij}^5} \; , \qquad
U_{ij}^{\al\bt} = \frac{3C_d}{a_{ij}^5} \left ( \dlt_{\al\bt} -
\; \frac{5a_{ij}^\al a_{ij}^\bt}{a_{ij}^2} \right ) \;  .
$$
The shift of the effective interactions, due to phonon excitations, according
to Eq. (94), reads as
\be
\label{102}
 \Dlt U_{ij} = \frac{3(5-d)}{d} \; r_0^2 \; \frac{C_D}{a_{ij}^5} \;  .
\ee
Taking into account only the nearest neighbors, yields
\be
\label{103}
\Dlt U_{ij} = \frac{3(5-d)}{d^2} \left ( \frac{r_0}{a} \right )^2 U_{ij} \;   .
\ee
Again, we see that the presence of phonons increases effective interactions.
This increase is not large, since, for $r_0 \ll a$, Eq. (103) shows that 
the shift is much smaller than $U_{ij}$.

For a well localized insulating state, the interaction matrix (30) leads to
\be
\label{104}
  \Phi_{ij}^{\al\bt} = 3 \; \frac{U_{ij}}{a_{ij}^2} \left ( \dlt_{\al\bt} -
\; \frac{5a_{ij}^\al a_{ij}^\bt}{a_{ij}^2} \right ) \; ,
\ee
which defines the dynamical matrix (56) as
\be
\label{105}
 D_{ij} = \frac{3(5-d)}{d} \; \frac{U_{ij}}{a_{ij}^2} \;  .
\ee
Taking into account only the nearest neighbors gives
\be
\label{106}
 D_0 =   \frac{3(5-d)}{d^{7/2}} \;  \frac{C_D}{a^5} \; .
\ee
The sound velocity (60) becomes
$$
 c_0 = \left [ \frac{3(5-d) C_D}{m a^3 d^{7/2}} \right ]^{1/2} \;  ,
$$
which for the Debye temperature (69) yields
\be
\label{107}
T_D = \frac{2\sqrt{3\pi(5-d)}}{d^{7/2}} \; \left [
\frac{d}{2}\; \Gm \left ( \frac{d}{2} \right ) \right ]^{1/d}
\sqrt{ \frac{C_D}{ma^5} } \;  .
\ee
In particular, for two and three dimensions, we find
$$
T_D = 3.162\; \sqrt{ \frac{C_D}{ma^5} } \qquad ( d = 2 ) \; ,
$$
\be
\label{108}
 T_D = 1.396\; \sqrt{ \frac{C_D}{ma^5} } \qquad ( d = 3 ) \;  .
\ee

It is convenient to introduce the {\it dipole length}
\be
\label{109}
 a_D \equiv \frac{mC_D}{\hbar^2} \;  ,
\ee
which sometimes is also called the effective dipole scattering length. Then
the Lindemann criterion (61) for the stability of a localized state can 
be written, depending on dimensionality, as
$$
\frac{a_D}{a} > 0.4 \qquad ( d = 2 ) \; ,
$$
\be
\label{110}
 \frac{a_D}{a} > 2.6 \qquad ( d = 3 ) \;  .
\ee
This tells us that the effective atomic interaction, i.e., the dipole length,
has to be sufficiently strong for the stability of the localized state.

In order to estimate the characteristic interaction parameters, let us consider
the systems of cold trapped atoms of $^{52}$Cr \cite{Griesmaier_23}, $^{168}$Er
\cite{Aikawa_24}, and $^{164}$Dy \cite{Lu_25}. We keep in mind that the typical
nearest-neighbor distance in an optical lattice is $a \sim 10^{-5}$ cm.

The dipole magnetic moment of $^{52}$Cr is $\mu_0 = 6 \mu_B$, which gives
$\mu_0 = 5.564 \times 10^{-20}$ erg/G. The atomic mass is
$m = 0.863 \times 10^{-22}$ g. The scattering length is $a_s = 103$ $a_B$.
The sound velocity is $c_0 \sim 0.2$ cm/s. The Debye frequency is
$\omega_D \sim 0.6 \times 10^5$ 1/s. The Debye temperature is
$T_D \sim 0.7 \times 10^{-22}$ erg or $T_D \sim 0.5 \times 10^{-6}$ K.
Taking $C_D \sim \mu_0^2$, we get $a_D = 2.4 \times 10^{-7}$ cm. This is much
shorter than the typical lattice distance $a$. That is, the stability
criterion (110) cannot be satisfied.

For $^{168}$Er, the dipole moment is $\mu_0 = 7 \mu_B$, which yields
$\mu_0 = 6.492 \times 10^{-20}$ erg/G. The mass is $m = 2.777 \times 10^{-22}$ g.
Then the dipole length is $a_D = 1.052 \times 10^{-6}$ cm. This is also shorter
than the typical intersite distance $a$. Hence the stability criterion (110)
is again not valid.

In the case of $^{164}$Dy, the dipole moment is $\mu_0 = 10 \mu_B$, which
results in $\mu_0 = 9.274 \times 10^{-20}$ erg/G. The mass is
$m = 2.698 \times 10^{-22}$ g. This gives the dipole length
$a_D = 2.087 \times 10^{-6}$ cm that again does not satisfy the stability
criterion (110).

Thus, phonon excitations do not allow for the formation of localized states
for the above atoms with pure dipolar forces. But there exist polar molecules
for which magnetic (or electric) moments can reach $100 \mu_B$ \cite{Pupillo_7}.
Such polar molecules, with the dipolar lengths several orders larger than those
of the above atoms, can satisfy the stability criterion, hence, can form
localized states that are stable against phonon collective excitations.
And even when absolutely stable states are not allowed, there can exist very 
long-lived metastable states.

\section{Conclusion}

We have studied the influence of phonon collective excitations on the 
possibility of cold atoms to form localized states in optical lattices. 
The appearing phonon excitations are self-organized atomic fluctuations,
caused by intersite atomic interactions collectivizing the atoms. 

It turns out that such phonon excitations are very important even for 
atoms with local delta-function interactions. Phonon oscillations can 
destroy the insulating state that would exist without them. 

The physical mechanism, by which phonons can destabilize an insulating state,
is the fact that phonons decrease the system energy by increasing the hopping
parameter.

The localized state can be stabilized by strong atomic interactions. These 
conclusions are valid for both local as well as long-range dipolar interactions. 
Taking into account collective phonon excitations is necessary when studying 
whether cold atoms in optical lattices can form localized insulating states 
in equilibrium. Even when in absolute equilibrium a localized state cannot 
exist, being unstable with respect to phonon excitations, this does not 
prohibit the existence of metastable localized states in optical lattices,
which live so long that they can be easily observed and studied. 

The reason why the conditions for the existence of localized states for a pure
Hubbard model without phonons and that one with the phonon excitations are
different is easy to understand. In a pure Hubbard model, the state is localized
when the width $l_0$ of the wave packet at a lattice site is narrow, such that
it is much smaller than the intersite distance $a$, or when $k_0 l_0 \ll 1$.
The latter, since $k_0 l_0 = (E_R/V_0)^{1/4}$, implies that $V_0 \gg E_R$.
Exactly this condition is observed in the experiments studying the insulating
Mott state in optical lattices \cite{Morsh_1,Moseley_2,Bloch_3,Yukalov_4,Greiner_23}.

However, the wave packet can be narrow, but strongly oscillating around a
lattice site. Such oscillations are characterized by the mean-square deviation
$r_0$. Then the localized state can exist if these oscillating wave packets
do not intersect with each other, which is the meaning of the Lindemann
criterion of stability, requiring that $r_0/a$ be at least smaller than $1$.
The latter imposes the condition that atomic interactions be sufficiently
strong, satisfying inequalities (89) or (90), depending on the space 
dimensionality. The condition $l_0 \ll a$ does not directly involve the
parameters of intersite atomic interactions, while the Lindemann criterion
of instability strongly depends on the parameters of such interactions. This
is why these criteria are principally different and do not need to coincide.

\vskip 2mm

{\bf Acknowledgment}

\vskip 2mm

One of the authors (V.I.Y.) acknowledges financial support from the Augsburg 
University and from the Russian Foundation for Basic Research 
(grant 14-02-00723).

\newpage

{\bf Appendix A. Hopping term and intersite interactions}

\vskip 2mm

In deriving the extended Hubbard Hamiltonian, one employs the basis of Wannier 
functions. It is known that this basis is convenient because Wannier functions 
can be made well localized \cite{Marzari_32,Souza_33,Modugno_33}, such that 
the tight-binding approximation becomes applicable, when the Wannier functions
are close to harmonic wave packets, 
$$
w(\br) = \prod_{\al=1}^d w_\al(\br) \; , \qquad
w_\al(\br) = \left ( \frac{m \om_\al}{\pi} \right )^{1/4}
\exp \left ( -\; \frac{m}{2} \; \om_\al r_\al^2 \right ) \; .
$$
The wave packets are well localized in the sense that
$$
 \frac{l_\al}{a_\al}  \ll 1 \qquad
\left ( l_\al \equiv \frac{1}{\sqrt{m\om_\al} } \right ) \; ,
$$
or, in other words, that
$$
 k_0^\al l_\al \ll 1 \qquad ( k_0^\al a_\al = \pi ) \;  .
$$

The value $l_\alpha$ plays the role of the localization length, or width,
for the wave packet. The corresponding oscillator frequency $\omega_\alpha$
can be found by minimizing the energy of the whole atomic system
\cite{Yukalov_4}, which takes into account atomic interactions. When the
packet width is mainly defined by the optical potential, then
$$
 \omega_\al = \sqrt{ \frac{2}{m} \; V_\al } \;
k_0^\al \qquad \left ( k_0^\al \equiv \frac{\pi}{a_\al} \right ) \;  .
$$

In this approximation, the hopping term (6) is
$$
  J_{ij} = \sum_{\al=1}^d \left [
\frac{\om_\al}{8} \left ( \frac{a_{ij}^\al}{l_\al} \right )^2 - V_\al \right ]
\exp \left \{ - \; \frac{1}{4}
\sum_{\al=1}^d \left ( \frac{a_{ij}^\al}{l_\al} \right )^2 \right \}
$$
and the matrix element (14) becomes
$$
 U_{ij}^{loc} = \frac{\Phi_d}{(2\pi)^{d/2} } \left (
\prod_{\al=1}^d \frac{1}{l_\al} \right ) \exp \left \{ - \; \frac{1}{2}
\sum_{\al=1}^d \left ( \frac{a_{ij}^\al}{l_\al} \right )^2 \right \} ,
$$
with the notation
$$
\ba_{ij} \equiv \ba_i - \ba_j = \{ a_{ij}^\al \} \; , \qquad
a_{ij}^2 \equiv | \ba_{ij} |^2 = \sum_{\al=1}^d \left ( a_{ij}^\al \right )^2 \; .
$$

For a cubic lattice, with $V_\alpha = V_0$, $\omega_\alpha = \omega_0$, and
$l_\alpha = l_0$, the hopping term simplifies to
$$
 J_{ij} = \left ( \frac{\om_0 a_{ij}^2}{8 l_0^2 }  - V_0 d \right )
\exp \left ( -\; \frac{a_{ij}^2}{4l_0^2} \right ) \;  ,
$$
while the above intersite interaction becomes
$$
U_{ij}^{loc} = U_{loc} \exp \left ( -\; \frac{a_{ij}^2}{2l_0^2} \right ) \;   ,
$$
with the on-site interaction, due to the local potential, being
$$
 U_{loc} =  \frac{\Phi_d}{(2\pi)^{d/2}l_0^d} \; .
$$

For a cubic lattice, with the frequency $\omega_0$ defined by the optical
potential, we have
$$
 \om_0 = 2 \sqrt{V_0 E_R } = \frac{\pi}{a} \sqrt{\frac{2}{m} \; V_0 }
$$
and
$$
 l_0 \equiv \frac{1}{\sqrt{m\om_0}} = \frac{1}{(4m^2 V_0 E_R)^{1/4} } \;  ,
$$
with $E_R = \pi^2 / 2 m a^2$. In the case of the nearest neighbors, using the
relations
$$
 \frac{\om_0 a_{ij}^2}{8 l_0^2 } = \frac{\pi^2}{4} \; V_0 d \; , \qquad
l_0^2 = \frac{a}{\pi\sqrt{2mV_0} } \; , \qquad
\frac{a^2}{l_0^2 } = \pi^2 \sqrt{\frac{V_0}{E_R} } \; ,
$$
we find the hopping term
$$
J = V_0 d \left ( \frac{\pi^2}{4} - 1 \right )
\exp \left ( - \; \frac{a^2 d}{4 l_0^2} \right )
$$
and the on-site interaction, caused by the local potential,
$$
U_{loc} = \frac{\Phi_d}{\pi^{d/2}} ( m^2 V_0 E_R ) ^{d/4} =
\frac{2\om_0 a_{eff}}{\sqrt{2\pi}\;l_\perp^{3-d} l_0^{d-2}} \; .
$$

The ratio of the on-site interaction to the hopping term reads as
$$
  \frac{U_{loc}}{J} =
\frac{4\Phi_d \exp(a^2d/4l_0^2)}{(\pi^2-4)(2\pi)^{d/2}\; l_0^d\; V_0 d} \; ,
$$
or it may be presented as
$$
  \frac{U_{loc}}{J} = \frac{4\Phi_d}{(\pi^2-4)d} \left (
\frac{m}{2a^2 V_0^3} \right )^{d/4}
\exp \left ( \frac{\pi d}{4} \sqrt{2mV_0 } \right ) \;   .
$$
This shows that, for a sufficiently deep lattice and large scattering length,
the value of $U_{loc}$ can be made much larger than $J$, so that the system
would be in a well localized insulating state.

\newpage

{\bf Appendix B. Mean-field illustration of phonon instability}

\vskip 2mm

The conventional approach to determine the properties of both, the phonons and
atoms, is based on a self-consistent evaluation of the self energy
(Migdal approximation) \cite{abrikosov63}. The latter provides an effective
(or renormalized) energy and its imaginary part, an effective scattering rate.
Such a static approximation might be insufficient, since it does not take into
account thermal fluctuations. It is possible to treat quantum and thermal
fluctuations separately. To this end, we can replace the phonon operators $b_{ks}$,
$b^\dagger_{ks}$ in (41) by their quantum average:
$b_{ks}\approx \langle b_{ks}\rangle\equiv v_{ks}$ and
$b_{ks}^\dagger\approx \langle b^\dagger_{ks}\rangle\equiv v_{ks}^*$.
In this approximation, we can keep thermal fluctuations but ignore quantum
fluctuations of the phonons. The atoms, on the other hand, are studied in full
quantum dynamics. This reduces the grand-canonical ensemble at inverse temperature
$\beta\equiv 1/k_BT$, defined by the generating  function $Tr e^{-\beta H}$, to
a functional integral with respect to thermal fluctuations of the lattice 
distortions ${\bf u}_j$ and a trace with respect to the quantum states of the 
atoms \cite{ziegler05}.

In order to illustrate how the phonon instability can develop, we consider 
a simplified model to show how phonon fluctuations rise with temperature. For 
this purpose we choose as the effective quasi-atom Hamiltonian $h$ a hopping 
term and a term that describes the displacement of atoms on nearest-neighbor 
sites. Assuming a bipartite lattice, the atomic Hamiltonian ${\hat H}_{at}$ 
of Eq. (35) can be reduced to the effective form
$$
h=\begin{pmatrix}
\mu & h_1-ih_2 \cr
h_1+ih_2 & \mu \cr
\end{pmatrix}
\ , \ \ h_1=({\hat H}_{at}+{\hat H}_{at}^T)/2, \ h_2=i({\hat H}_{at}-{\hat H}_{at}^T)/2
\ ,
$$
where $\mu$ is an effective chemical potential of atoms. Within the mean-field 
approximation, this leads to a spatially uniform displacement field $u$ entering
the action
$$
S=\beta {\hat H}_{vib}+\int\ln\left[1+e^{-2\beta\mu}-2e^{-\beta\mu}
\cosh\left(\beta\sqrt{h_1^2+h_2^2}\right)\right]\frac{d{\bf k}}{(2\pi)^d} \; , 
$$
with the phonon Hamiltonian of Eq. (38). The calculation of the phonon free 
energy $-\beta\ln Z$ requires the integration over the atomic displacement field $u$, 
which can be performed in saddle-point approximation fixing the displacement field 
by the saddle-point equation
$$
0=\frac{\partial S}{\partial u^\alpha}=\beta\frac{\partial {\hat H}_{vib}}{\partial u^\alpha}
-\int\frac{\sinh\left(\beta\sqrt{h_1^2+h_2^2}\right)}
{\cosh(\beta\mu)-\cosh\left(\beta\sqrt{h_1^2+h_2^2}\right)}
\frac{h_1\partial h_1/\partial u^\alpha+h_2\partial h_2/\partial u^\alpha}{\sqrt{h_1^2+h_2^2}}
\frac{d{\bf k}}{(2\pi)^d}
\ .
$$
Phonon fluctuations around the saddle point are described by the fluctuation matrix
\[
\frac{\partial^2 S}{\partial u^\alpha \partial u^\beta}
=\beta\frac{\partial^2 {\hat H}_{vib}}{\partial u^\alpha \partial u^\beta}
-\beta^2\int\frac{\cosh\left(\beta\sqrt{h_1^2+h_2^2}\right)\cosh(\beta\mu)-1}
{\left[\cosh(\beta\mu)-\cosh\left(\beta\sqrt{h_1^2+h_2^2}\right)\right]^2}
\]
$$
\times
\frac{(h_1\partial h_1/\partial u^\alpha+h_2\partial h_2/\partial u^\alpha)
(h_1\partial h_1/\partial u^\beta+h_2\partial h_2/\partial u^\beta)}{h_1^2+h_2^2}
\frac{d{\bf k}}{(2\pi)^d}
\ .
$$
The effect of the phonon fluctuations increases with decreasing eigenvalues of 
the fluctuation matrix. The negative sign in front of the positive integral 
reflects the fact that the phonon fluctuations increase with increasing temperature 
and can eventually lead to an instability of the atomic system. Thus, the 
fluctuation matrix is related to the Lindemann criterion of Sect. \ref{sect:inst}. 
This behavior is depicted by the temperature dependence of the integrand in Fig. 1. Eigenvalues can even become negative, which indicates a phase transition. The 
latter could either be a structural phase transition of the lattice 
\cite{ziegler11,Ziegler_29} or a melting transition \cite{Lindemann_23}.

\newpage

\begin{center}
{\large {\bf Figure Caption} }
\end{center}

\vskip 2cm

The integrand of the last equation of Appendix B as a function of temperature $T$ 
(in units of the hopping energy), describing the increase of phonon fluctuations 
with rising temperature, for $\sqrt{h_1^2+h_2^2}=0.01,0.8,1$ (from bottom to top),
$d = 3$, and $\mu = 1.2$.

\newpage

\begin{figure*}[t]
\begin{center}
\includegraphics[width=9cm]{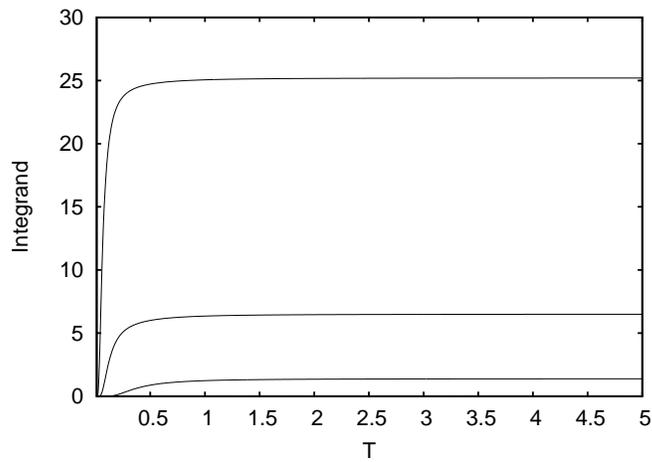}
\caption{The integrand of the last equation of Appendix B as a function of 
temperature $T$ (in units of the hopping energy), describing the increase of 
phonon fluctuations with rising temperature, for 
$\sqrt{h_1^2+h_2^2}=0.01,0.8,1$ (from bottom to top),
$d = 3$, and $\mu=1.2$.
}
\end{center}
\label{fig:fluct}
\end{figure*}

\end{document}